\documentclass[reprint,prd,12pt,superscriptaddress,onecolumn,nofootinbib,aps]{revtex4-2}
\usepackage{hyperref}
\usepackage{ifpdf}
\usepackage{slashed}
\usepackage{subfig}
\usepackage[normalem]{ulem}
\usepackage{amsmath,amssymb,amsfonts,gensymb}
\usepackage{rotating}
\usepackage{nicefrac}
\usepackage{color,xcolor}
\usepackage{fancyhdr}
\usepackage{lineno}
\usepackage{braket}
\usepackage{multirow,adjustbox}
\usepackage{empheq}
\usepackage{graphics}
\usepackage{mathtools}

\definecolor{bv}{rgb}{0.54, 0.17, 0.89}
\allowdisplaybreaks

\def\bar {\overline}

\def\beq{\begin{equation}}
\def\eeq{\end{equation}}
\def\bea{\begin{eqnarray}}
\def\eea{\end{eqnarray}}
\def\barr{\begin{array}}
\def\earr{\end{array}}

\usepackage{color}

\usepackage[normalem]{ulem}

\definecolor{darkgreen}{cmyk}{1,0,1,0.4}

\def\com2#1{\textcolor{red}{\textit{#1}}}

\begin{document}

\author{Shibasis Roy}	
\email{shibasis.cmi@gmail.com}
\affiliation{Chennai Mathematical Institute, Siruseri 603103, Tamil Nadu, India}

\title{$CP$-violation and $U$-spin symmetry in four-body bottom baryon decays}

\begin{abstract}
In view of the recent observation of $CP$-violation in bottom baryon decays, we explore $CP$-asymmetry relations in charmless four-body decays of bottom baryons featuring fully charged hadronic final states. Starting with a general effective Hamiltonian, we derive amplitude relations among various decay modes within the framework of $U$-spin symmetry. These relations lead to symmetry-based predictions for $CP$-violating rate asymmetries in $U$-spin conjugate decay channels. Furthermore, we demonstrate that all leading-order amplitude relations are violated once $U$-spin breaking effects are taken into account.

\end{abstract}

\maketitle

\section{Introduction}
The discovery of $CP$ violation (CPV) in bottom baryon decays at LHCb~\cite{LHCb:2025ray} heralds a new era in experimental observations of $CP$ violation within the Standard Model. Since the observed matter-antimatter asymmetry~\cite{Sakharov:1967dj} of the universe is closely tied to baryons, which make up the visible universe, $CP$ violation in baryons provides a direct connection to this phenomenon. While $CP$ violation in meson mixing and decays has been well-established through experiments~\cite{Christenson:1964fg,BaBar:2001ags,Belle:2001zzw,CDF:2011ubb,CDF:2014pzb,LHCb:2019hro,Belle:2023str}, the status of CPV in baryons remained uncertain, despite significant efforts from LHCb~\cite{LHCb:2016yco,LHCb:2018fpt,LHCb:2018fly,LHCb:2019jyj,LHCb:2019oke,LHCb:2021enr,LHCb:2024iis,LHCb:2025bfy} in recent years. In conjunction with the evidence of CPV in the $\Lambda^{0}_{b} \to \Lambda K^{+}K^{-}$ decay~\cite{LHCb:2024yzj}, the measured $CP$ asymmetry in the $\Lambda^{0}_{b} \to p^{+}K^{-}\pi^{+}\pi^{-}$ decay~\cite{LHCb:2025ray},
\begin{align}
    A_{CP}(\Lambda^{0}_{b}\to p^{+}K^{-}\pi^{+}\pi^{-})=(2.45\pm 0.46 \pm 0.10)\%
\end{align}
represents a seminal result in our understanding of the origin of $CP$ violation within the Standard Model. Since CPV in both mesons and baryons arises from the CKM mechanism~\cite{Cabibbo:1963yz,Kobayashi:1973fv} that governs quark transitions, a crucial test of the theory comes from flavor symmetries, particularly given the challenges in making precise theoretical predictions for $CP$ violation in multibody baryon decays. Several approaches have been proposed to study $CP$ violation in multibody bottom baryon decays~\cite{Gronau:2015gha, Durieux:2015zwa,Durieux:2016nqr}, including a method to extract the $CP$-violating weak phase in charmless bottom baryon decays~\cite{Sinha:2021mmx,Roy:2023gcl}.

In this paper, we use a subclass of flavor symmetry, known as the $U$-spin symmetry~\cite{Gell-Mann:1964ewy,Zweig:1964jf}, reflecting the symmetry between down type quarks $d$ and $s$ to relate $b\to d$ and $b\to s$ transition amplitudes~\cite{Gronau:2000zy,Gronau:2013mda}. In our approach, we first construct the $U$-spin decomposition~\cite{Gronau:2000zy,Grossman:2012ry,Gronau:2013mda,Grossman:2018ptn} of the most general Hamiltonian mediating a transition from the bottom baryon to a four body charmless final state. In the next step, we consider the dimension-6 effective Hamiltonian of the SM that mediates such transition and map its $U$-spin components to the corresponding components of the general Hamiltonian. At the leading-order $U$-spin limit, a set of amplitude relations among the four-body final states accessible to the bottom baryons naturally emerges. Moreover, this method allows us to extend the analysis beyond the exact $U$-spin symmetry limit and determine whether specific leading-order amplitude relations continue to hold. This feature is particularly significant given that $U$-spin symmetry is broken by the mass difference between the $d$ and $s$ quarks, with the breaking estimated to be at the level of $\sim$ 30\%~\cite{Gronau:2013mda}. We wish to emphasize, however, that our $U$-spin symmetry–based argument is intended primarily as a guiding framework within a data-driven approach, useful for estimating decay amplitudes and identifying promising four-body bottom baryon decays where sizable $CP$ violation may be observed. 

There are several $U$-spin based analysis~\cite{Wang:2024rwf,He:2025msg,Zhang:2025jnw,Chen:2025drl} available that connects the recently measured CPV observed in $\Lambda^{0}_{b}$ decay with their corresponding $U$-spin partners under specific assumptions. Our approach differs from previous analyses by first enumerating the maximum number of invariant $U$-spin amplitudes permitted in the $B_{b} \to B_{1}PP'P_{1}$ transition, where $B_{b}$ is the initial bottom baryon, $P$ and $P'$ are pseudoscalar mesons with the same electric charge, and $P_{1}$ is a pseudoscalar meson with opposite charge, and then systematically relating or neglecting these amplitudes based on the $U$-spin decomposition of the Standard Model dimension-6 effective Hamiltonian. It is worth highlighting that our approach closely mirrors the method first introduced in~\cite{Grinstein:1996us} to study $B$-meson decays using $SU(3)$ flavor symmetry, and subsequently applied to two-body hadronic decays of bottom baryons in~\cite{Roy:2019cky,Roy:2020nyx}.

In Section~\eqref{Sec-II}, we discuss the theoretical framework for performing a $U$-spin decomposition of bottom baryon decay amplitudes involving fully charged four-body final states. Section~\eqref{sec: amp rel} focuses on derivation of symmetry-based relations between physical observables, such as $CP$-violating rate asymmetries. We conclude by commenting on the effect of $U$-spin breaking on these observables.

\section{Application of $U$-spin to decay amplitudes}
\label{Sec-II}
Under $U$-spin, the final state charged pseudoscalar mesons form doublets of $U$-spin~\cite{Lichtenberg:1978pc,Soni:2006vi} expressed as, \begin{align}
    P, \, P'=\begin{pmatrix}
        \pi^{-}\\
        K^{-}
    \end{pmatrix}
    = \begin{pmatrix}
        -\vert \frac{1}{2}, \frac{1}{2}\rangle\\
        -\vert \frac{1}{2}, -\frac{1}{2}\rangle
    \end{pmatrix},
    \quad
    \begin{pmatrix}
        K^{+}\\
        \pi^{+}
    \end{pmatrix}
    =\begin{pmatrix}
        \vert \frac{1}{2}, \frac{1}{2}\rangle\\
        -\vert \frac{1}{2}, -\frac{1}{2}\rangle
    \end{pmatrix}
\end{align}
Similarly, a subset of the light baryons belonging to the octet under $SU(3)$-flavor form distinct doublets under $U$-spin given below,
\begin{align}
    B_{1}^{+}=\begin{pmatrix}
        p^{+} \\
        \Sigma^{+}
    \end{pmatrix}=\begin{pmatrix}
        \vert \frac{1}{2}, \frac{1}{2}\rangle\\
        \vert \frac{1}{2}, -\frac{1}{2}\rangle
    \end{pmatrix}, \quad
B_{1}^{-}=\begin{pmatrix}
        \Sigma^{-} \\
        \Xi^{-}
    \end{pmatrix}=\begin{pmatrix}
        \vert \frac{1}{2}, \frac{1}{2}\rangle\\
        \vert \frac{1}{2}, -\frac{1}{2}\rangle
    \end{pmatrix}.
\end{align}
The bottom baryon under $SU(3)$-flavor symmetry transforms as an anti-triplet while its neutral components $\Lambda^{0}_{b}$ and $\Xi_{b}^{0}$ form an $U$-spin doublet $B_{b}=\begin{pmatrix}
    \Lambda^{0}_{b} & \Xi_{b}^{0}.
\end{pmatrix}$
Given the initial state ($\mathbf{i}$) and final states ($\mathbf{f}$), we can write down the most general transition connecting the initial and final state symbolically as,
\begin{align}
    \mathcal{H}_{\text{eff}}\equiv \mathbf{f}\otimes \bar{\mathbf{i}}.
\end{align}
For the final state we use the order $(P\otimes P^{'})\otimes P_{1}\otimes B_{1}$ to calculate the tensor product and subsequently apply the Wigner-Eckart theorem
for $U$-spin decomposition. The final state $\mathbf{f}$ made off two identically charged mesons, a third oppositely charged meson and a light baryon must be symmetrized over the identically charged mesons tensor product state as they are indistinguishable under exact $U$-spin~\cite{Sinha:2021mmx}. Operationally we have,
\begin{align}
    \{P^{\pm} P^{'\pm}  \}=\frac{1}{\sqrt{2}}(P^{\pm} P'^{\pm}+ P^{'\pm} P^{\pm})
\end{align}
We highlight that this symmetrization process misses an additional $\sqrt{2}$ factor whenever two identical mesons $P^{\pm}(U,U_{3})=P^{'\pm}(U^{'}=U, U_{3}^{'}=U_{3})$ are encountered which we account for in our calculation. Now, we can perform the $U$-spin decomposition of the decay amplitude as,
\pagebreak
\begin{widetext}
\vspace{-0.2cm}
	\begin{eqnarray}
	\label{Master Formula}
	\mathcal{A}(B_{b} \to B_{1}^{\pm}PP'P_{1})=\quad\sum^{U^{P}+U^{P'},U^{M^{0}}+U^{P_{1}}\dots}_{\mathclap 
	{\substack{\{U^{M^{0}}=\vert U^{P}-U^{P'}\vert,	\, U^{M}=\vert U^{M^{0}}-U^{P_{1}}\vert, \, U^{F} \dots \} \\
				U^{P}_{3}+U^{P'}_{3}=U_{3}^{M_{0}}, \, U^{M_{0}}_{3}+ U_{3}^{P_{1}}=U_{3}^{M}\\
				U_{3}^{M} + U_{3}^{B_{1}}=U_{3}^{F},\, 
				U_{3}^{F} -U_{3}^{B_{b}}= U_{3}^{\mathcal{H}}}}} 
	C^{\{U^{P}_{3} U^{P'}_{3} U_{3}^{M_{0}}\}}_{U^{P} U^{P'} U^{M_{0}}}
    C^{U_{3}^{M_{0}}\, U_{3}^{P_{1}}\, U_{3}^{M} }_{U^{M^{0}}\, U^{P_{1}}\, U^{M}} 
    \\\times C^{ U_{3}^{M} \, U_{3}^{B_{1}}\, U_{3}^{F} }_{U^{M}\, U^{B_{1}}\, U^{F}} C^{U_{3}^{F}\, (-U_{3}^{B_{b}})\, U_{3}^{\mathcal{H}} }_{U^{F}\, U^{B_{b}}\, 
	U^{\mathcal{H}}} 
	A^{U^{M},U^{F},\pm}(U^{\mathcal{H}})\nonumber
	\end{eqnarray}     
\end{widetext}
where $A^{i,j,\pm}(U^\mathcal{H})$ denotes the $U$-spin subamplitudes and $C^{U_{3}^{a} U_{3}^{b} U_{3}^{c}}_{U^{a} U^{b} U^{c}}$ represents the $SU(2)$ Clebsch–Gordan coefficients. The $\pm$ label in $A^{i,j,\pm}(U^\mathcal{H})$ distinguishes the $U$-spin doublet to which the final state light baryon belongs, and in general $A^{i,j,+}(U^\mathcal{H})$ and $A^{i,j,-}(U^\mathcal{H})$ are unrelated. The subamplitudes $A^{i,j,\pm}(U^\mathcal{H})$ in the most general case also involve unknown dynamical coefficients, for example, in the form of Wilson coefficients and CKM factors, that only get fixed when a particular form of the effective Hamiltonian is assumed. Therefore, apriori $A^{i,j,\pm}(U^\mathcal{H})$ are all independent of each other. We now proceed to a discussion of the lowest order effective Hamiltonian~\cite{Ciuchini:1993vr,Buchalla:1995vs,Buras:1998raa}
mediating charmless bottom baryon $\Delta \vert S \vert=1$ and $ \Delta S=0$ transitions. Both parts of the effective Hamiltonian is composed of the 
operators $ O_{1}$,\ldots, $ O_{10}$ and can be written 
as:
\begin{widetext}
\begin{align}
\label{eff H}
\mathcal{H}_{\text{eff}}=\frac{4G_{F}}{\sqrt{2}}\Big[\lambda^{(s)}_{u}
\Big(C_{1}(O^{(u)}_{1}-O^{(c)}_{1})+C_{2}(O^{(u)}_{2}-O^{(c)}_{2})\Big)-
\lambda^{(s)}_{t} \sum_{i=1,2}C_{i}O^{(c)}_{i}-
\lambda^{(s)}_{t}
\sum_{i=3}^{10}C_{i}O_{i}^{(s)}\nonumber \\ +\lambda^{(d)}_{u}
\Big(C_{1}(O^{(u)}_{1}-O^{(c)}_{1})+C_{2}(O^{(u)}_{2}-O^{(c)}_{2})\Big)-\lambda^{(d)}_{t} \sum_{i=1,2}C_{i}O^{(c)}_{i}-
\lambda^{(d)}_{t}
\sum_{i=3}^{10}C_{i}O_{i}^{(d)} \Big],
\end{align} 
\end{widetext}
where $V_{ub}V_{us}^{*}=\lambda_{u}^{s}$, $V_{ub}V_{ud}^{*}=\lambda_{u}^{d}$, 
$V_{tb}V_{ts}^{*}=\lambda_{t}^{s}$, $V_{tb}V_{td}^{*}=\lambda_{t}^{d}$ are the 
CKM elements and $C_{i}$ s are the Wilson coefficients. $O_{1}$ and $ O_{2}$ 
are the ``tree'' operators:
\begin{align}
O_{1}^{(u)} & =  (\overline{u}^{i}_{L}\gamma^{\mu}b^{j}_{L})
(\overline{s}^{j}_{L}\gamma_{\mu}u^{i}_{L}) \nonumber \\
O_{1}^{(c)} & =  (\overline{c}^{i}_{L}\gamma^{\mu}b^{j}_{L})
(\overline{s}^{j}_{L}\gamma_{\mu}c^{i}_{L}) \nonumber\\
O_{2}^{(u)}&=(\overline{u}^{i}_{L}\gamma^{\mu}b^{i}_{L})
(\overline{s}^{j}_{L}\gamma_{\mu}u^{j}_{L})  \nonumber  \\
O_{2}^{(c)}&=(\overline{c}^{i}_{L}\gamma^{\mu}b^{i}_{L})
(\overline{s}^{j}_{L}\gamma_{\mu}c^{j}_{L}). 
\end{align}
$ O_{3}$, \ldots, $ O_{6}$ are the ``gluonic penguin'' operators:
\begin{align}
O_{3}^{(s)} & = (\overline{s}^{i}_{L}\gamma^{\mu}b^{i}_{L})
\sum_{q=u,d,s}(\overline{q}^{j}_{L}\gamma_{\mu}q^{j}_{L})\nonumber \\
O_{4}^{(s)} & = (\overline{s}^{i}_{L}\gamma^{\mu}b^{j}_{L}) 
\sum_{q=u,d,s}(\overline{q}^{j}_{L}\gamma_{\mu}q^{i}_{L})\nonumber \\
O_{5}^{(s)} & = (\overline{s}^{i}_{L}\gamma^{\mu}b^{i}_{L})
\sum_{q=u,d,s}(\overline{q}^{j}_{R}\gamma_{\mu}q^{j}_{R})\nonumber \\
O_{6}^{(s)} & = (\overline{s}^{i}_{L}\gamma^{\mu}b^{j}_{L})
\sum_{q=u,d,s}(\overline{q}^{j}_{R}\gamma_{\mu}q^{i}_{R}). 
\end{align}
and finally $ O_{7}$, \ldots, $ O_{10}$ are the four ``Electroweak penguin" operators:  
\begin{align}
O_{7}^{(s)} & = \frac{3}{2} (\overline{s}^{i}_{L}\gamma^{\mu}b^{i}_{L})
\sum_{q=u,d,s}e_{q}(\overline{q}^{j}_{R}\gamma_{\mu}q^{j}_{R}),\nonumber \\ 
O_{8}^{(s)} & =\frac{3}{2} (\overline{s}^{i}_{L}\gamma^{\mu}b^{j}_{L})
\sum_{q=u,d,s}e_{q}(\overline{q}^{j}_{R}\gamma_{\mu}q^{i}_{R}),\\
O_{9}^{(s)} & = \frac{3}{2} (\overline{s}^{i}_{L}\gamma^{\mu}b^{i}_{L})
\sum_{q=u,d,s}e_{q}(\overline{q}^{j}_{L}\gamma_{\mu}q^{j}_{L}),\nonumber \\
O_{10}^{(s)} & =\frac{3}{2} (\overline{s}^{i}_{L}\gamma^{\mu}b^{j}_{L})
\sum_{q=u,d,s}e_{q}(\overline{q}^{j}_{L}\gamma_{\mu}q^{i}_{L}).
\end{align}
Restricting ourselves to a discussion of $U$-spin of these operators, we find that the tree operators $O_{1}$ and $O_{2}$ transform as a $U$-spin doublet with the up component corresponding to $b\to d$ transition ($\Delta S = 0$) while the down component corresponds to $b\to s$ transition ($ \Delta \vert S \vert = 1$) transition. The relevant CKM elements which form part of the dynamical coefficients are given, respectively, as $V_{ub}V_{ud}^{*}$ and $V_{ub} V_{us}^{*}$. Since the combination $d\bar{d}+s\bar{s}$ transforms as a $U$-spin singlet, the complete set of QCD and Electroweak penguin four-Fermi operators ($O_{3}\dots O_{10}$) transform as distinct $U$-spin doublets. Interestingly, unlike isospin, which is explicitly broken by electromagnetic interactions, leading to isospin-violating effects in the electroweak penguin operators, no such complication arises in the case of $U$-spin, since the $d$ and $s$ quarks carry the same electric charge. Therefore, to the leading order in $U$-spin, the dim-6 SM effective Hamiltonian only induces a doublet in Eq~\eqref{Master Formula}. We observe that certain higher dimensional $U$-spin representations present in the general Hamiltonian have vanishing dynamical coefficients (See Appendix~\eqref{App:1}) corresponding to non-existent $U$-spin subamplitudes.  
\section{Amplitude Relations}
\label{sec: amp rel}
Possible four body all-electrically charged final states originating from charmless bottom baryon decays are listed in Table~\eqref{Tab:1}.
\begin{table}[tb]
\centering
\setlength{\tabcolsep}{12pt}
\renewcommand*{\arraystretch}{1.3}	
\begin{tabular}{|c|c||c|c|}	
\hline \hline
			$\Delta \vert S \vert =1 $ modes & $\Delta  S  = 0 $ modes & $\Delta \vert S \vert =1 $ modes & $\Delta  S  = 0 $ modes  \\ 
			\hline
			$\Lambda^{0}_{b}\to p^{+}K^{-}\pi^{-}\pi^{+}$	& $\Xi^{0}_{b}\to \Sigma^{+}K^{-}\pi^{-}K^{+}$	& $\Xi^{0}_{b}\to p^{+}K^{-}K^{-}\pi^{+}$       & $\Lambda^{0}_{b}\to \Sigma^{+}\pi^{-}\pi^{-}K^{+}$		   \\
			\hline
			$\Lambda^{0}_{b}\to \Xi^{-}K^{+}K^{+}K^{-}$	    & $\Xi^{0}_{b}\to \Sigma^{-}\pi^{+}\pi^{+}\pi^{-}$	& $\Xi^{0}_{b}\to \Xi^{-}\pi^{+}K^{+}K^{-}$       & $\Lambda^{0}_{b}\to \Sigma^{-}K^{+}\pi^{+}\pi^{-}$			\\
			\hline
			$\Lambda^{0}_{b}\to \Sigma^{+}\pi^{-}\pi^{-}\pi^{+}$  &	$\Xi^{0}_{b}\to p^{+}K^{-}K^{-}K^{+}$	& $\Xi^{0}_{b}\to \Sigma^{+}K^{-}\pi^{-}\pi^{+}$   &   $\Lambda^{0}_{b}\to p^{+}\pi^{-}K^{-}K^{+}$	\\
			\hline
			$\Lambda^{0}_{b}\to \Sigma^{-}K^{+}\pi^{+}K^{-}$	& $\Xi^{0}_{b}\to \Xi^{-}\pi^{+}K^{+}\pi^{-}$	&  $\Xi^{0}_{b}\to \Sigma^{-}\pi^{+}\pi^{+}K^{-}$    & $\Lambda^{0}_{b}\to \Xi^{-}K^{+}K^{+}\pi^{-}$	    \\
            \hline
            $\Lambda^{0}_{b}\to p^{+}K^{-}K^{-}K^{+}$		&	$\Xi^{0}_{b}\to \Sigma^{+}\pi^{-}\pi^{-}\pi^{+}$	&  $\Xi^{0}_{b}\to \Xi^{-}\pi^{+}\pi^{+}\pi^{-}$    & $\Lambda^{0}_{b}\to \Sigma^{-}K^{+}K^{+}K^{-}$	    \\
            \hline
            $\Lambda^{0}_{b}\to \Xi^{-}K^{+}\pi^{+}\pi^{-}$	 &	$\Xi^{0}_{b}\to \Sigma^{-}\pi^{+}K^{+}K^{-}$ & $\Xi^{0}_{b}\to \Sigma^{+}K^{-}K^{-}K^{+}$    & $\Lambda^{0}_{b}\to p^{+}\pi^{-}\pi^{-}\pi^{+}$	   \\
            \hline
            $\Lambda^{0}_{b}\to \Sigma^{+}K^{-}\pi^{-}K^{+}$	& $\Xi^{0}_{b}\to p^{+}\pi^{-}K^{-}\pi^{+}$	&  $\Lambda^{0}_{b}\to \Sigma^{-}\pi^{+}\pi^{+}\pi^{-}$	& $\Xi^{0}_{b}\to \Xi^{-}K^{+}K^{+}K^{-}$	 \\

            \hline \hline
\end{tabular}
\caption{Charmless four body decays of bottom baryons involving all charged final states }
\label{Tab:1}
\end{table}
We choose to represent the $\Delta S=0$ and $\Delta \vert S \vert =1$ decay amplitudes and the complete set of $U$-spin subamplitudes in terms of column matrices $\mathcal{A}$ and $A$ respectively such that there is coefficient matrix ($\mathbb{C}$) that links both basis through the equation,
\begin{align}
    \mathcal{A}=\mathbb{C} A
\end{align}
The explicit form the matrix $\mathbb{C}$ is given in Appendix~\eqref{App:1}. We notice that there are in general 14 independent $U$-spin subamplitudes describing all 28 decays. Alternatively, these 28 decay modes can be grouped based on the charge of the final state baryon. In that case, 14 decay modes with a positively charged baryon  are described in terms of 7 $A^{i,j,+}(U^\mathcal{H})$ subamplitudes while the remaining 14 decay modes with a negatively charged baryon are described in terms of 7 $A^{i,j,-}(U^\mathcal{H})$ subamplitudes. At this stage, no simple amplitude relation between only two decay modes can be established. Now assuming the SM Hamiltonian expressed in Eq~\eqref{eff H}, which contributes to the $U$-spin subamplitudes $A^{\frac{1}{2},0,\pm}(U^{\mathcal{H}}=\frac{1}{2}), \,
     A^{\frac{1}{ 2},1,\pm}(U^{\mathcal{H}}=\frac{1}{2}) $ and $
       A^{\frac{3}{ 2},1,\pm}(U^{\mathcal{H}}=\frac{1}{2})$ but not to  $ A^{\frac{1}{ 2},1,\pm}(U^{\mathcal{H}}=\frac{3}{2}),\,
        A^{\frac{3}{ 2},1,\pm}(U^{\mathcal{H}}=\frac{3}{2}),\,
         A^{\frac{3}{ 2},2,\pm}(U^{\mathcal{H}}=\frac{3}{2}),\,
           A^{\frac{3}{ 2},2,\pm}(U^{\mathcal{H}}=\frac{5}{2})$
for both the tree and penguin operators, the number of independent $U$-spin subamplitudes can be reduced to 6, three for each class of decays with a positively or negatively charged baryon in the final state. We can further decompose the decay amplitudes in terms of tree ($\mathcal{A}_{T}$) and penguin ($\mathcal{A}_{P}$) parts by factoring out the CKM elements $\lambda^{s,d}
_{u,t}$ from the decay amplitude and write,
\begin{align}
    \mathcal{A}^{(s,d)}=\lambda_{u}^{s,d}\mathcal{A}_{T}+\lambda^{s,d}_{t}\mathcal{A}_{P}
\end{align}
where  $(s, d)$ denote the $\Delta \vert S \vert = 1$, $\Delta S=0$ process respectively. 
\begin{table}[th!]
\begin{adjustbox}{width=\columnwidth,keepaspectratio}
\centering
\setlength{\tabcolsep}{12pt}
\renewcommand*{\arraystretch}{1.1}	
\begin{tabular}{cc}	
\hline \hline
			$B_{b}\to B_{1}^{+}PP^{'}P_{1}$ modes & $B_{b}\to B_{1}^{-}PP^{'}P_{1}$ modes  \\ 
			\hline
			$\mathcal{A}_{T(P)}(\Lambda^{0}_{b}\to p^{+}K^{-}\pi^{-}\pi^{+})=\mathcal{A}_{T(P)}(\Xi^{0}_{b}\to \Sigma^{+}K^{-}\pi^{-}K^{+})$ &    $\mathcal{A}_{T(P)}(\Lambda^{0}_{b}\to \Sigma^{-}K^{+}\pi^{+}K^{-})=\mathcal{A}_{T(P)}(\Xi^{0}_{b}\to \Xi^{-}\pi^{+}K^{+}\pi^{-})$		   \\
			 \hline
             
			 $\mathcal{A}_{T(P)}(\Lambda^{0}_{b}\to \Sigma^{+}\pi^{-}\pi^{-}\pi^{+})=\mathcal{A}_{T(P)}(\Xi^{0}_{b}\to p^{+}K^{-}K^{-}K^{+})$  & $\mathcal{A}_{T(P)}(\Lambda^{0}_{b}\to \Xi^{-}K^{+}K^{+}K^{-})=\mathcal{A}_{T(P)}(\Xi^{0}_{b}\to \Sigma^{-}\pi^{+}\pi^{+}\pi^{-})$   		\\
			 \hline
			
    	$\mathcal{A}_{T(P)}(\Lambda^{0}_{b}\to p^{+}K^{-}K^{-}K^{+})=\mathcal{A}_{T(P)}(\Xi^{0}_{b}\to \Sigma^{+}\pi^{-}\pi^{-}\pi^{+})$  & $\mathcal{A}_{T(P)}(\Lambda^{0}_{b}\to \Sigma^{-}\pi^{+}\pi^{+}\pi^{-})=\mathcal{A}_{T(P)}(\Xi^{0}_{b}\to \Xi^{-}K^{+}K^{+}K^{-})$     \\
        \hline
        $\mathcal{A}_{T(P)}(\Lambda^{0}_{b}\to \Sigma^{+}K^{-}\pi^{-}K^{+})=\mathcal{A}_{T(P)}(\Xi^{0}_{b}\to p^{+}\pi^{-}K^{-}\pi^{+})$ & $\mathcal{A}_{T(P)}(\Lambda^{0}_{b}\to \Xi^{-}\pi^{+}K^{+}\pi^{-})=\mathcal{A}_{T(P)}(\Xi^{0}_{b}\to \Sigma^{-}K^{+}\pi^{+}K^{-})$\\
        \hline 

        $\mathcal{A}_{T(P)}(\Lambda^{0}_{b}\to p^{+}\pi^{-}\pi^{-}\pi^{+})=\mathcal{A}_{T(P)}(\Xi^{0}_{b}\to \Sigma^{+}K^{-}K^{-}K^{+})$ & $\mathcal{A}_{T(P)}(\Lambda^{0}_{b}\to \Sigma^{-}K^{+}K^{+}K^{-}) = \mathcal{A}_{T(P)}(\Xi^{0}_{b}\to \Xi^{-}\pi^{+}\pi^{+}\pi^{-})$\\
        \hline

        $\mathcal{A}_{T(P)}(\Lambda^{0}_{b}\to \Sigma^{+}\pi^{-}\pi^{-}K^{+})=\mathcal{A}_{T(P)}(\Xi^{0}_{b}\to p^{+}K^{-}K^{-}\pi^{+})$ & $\mathcal{A}_{T(P)}(\Lambda^{0}_{b}\to \Xi^{-}K^{+}K^{+}\pi^{-})=\mathcal{A}_{T(P)}(\Xi^{0}_{b}\to \Sigma^{-}\pi^{+}\pi^{+}K^{-})$\\
        \hline

        $\mathcal{A}_{T(P)}(\Lambda^{0}_{b}\to p^{+}\pi^{-}K^{-}K^{+})=\mathcal{A}_{T(P)}(\Xi^{0}_{b}\to \Sigma^{+}\pi^{-}K^{-}\pi^{+})$ & $\mathcal{A}_{T(P)}(\Lambda^{0}_{b}\to \Sigma^{-}\pi^{+}K^{+}\pi^{-})=\mathcal{A}_{T(P)}(\Xi^{0}_{b}\to \Xi^{-}\pi^{+}K^{+}K^{-})$\\
   \hline     
\hline
\end{tabular}
\end{adjustbox}
\caption{Amplitude $U$-spin relations between $\Delta \vert S \vert =1$ and $\Delta S=0$ four body bottom baryon decays, assuming the unbroken SM Hamiltonian. The subscript $T$ and $P$ denotes the tree and penguin part of the full decay amplitude respectively.}
\label{Tab:2}
\end{table}
We obtain the following amplitude relations connecting the $\Delta \vert S \vert=1$ and $\Delta S=0$ sectors given in Table~\eqref{Tab:2}.
$A_{CP}$ is defined subsequently as,
\begin{align}
A_{CP}=&\frac{\Gamma(B_{b} \to B_{1}PP'P_{1})-\Gamma(\bar{B_{b}} \to \bar{B_{1}}\bar{P}\bar{P'}\bar{P_{1}})}
	{\Gamma(B_{b} \to B_{1}PP'P_{1})+\Gamma(\bar{B_{b}} \to \bar{B_{1}}\bar{P}\bar{P'}\bar{P_{1}})}\nonumber\\ 			    
	=&\frac{-\,4\textbf{J}
\times\text{Im}\Big[\mathcal{A}^{*}_{T}(B_{b} \to B_{1}PP'P_{1})\mathcal{A}_{P}(B_{b} \to B_{1}PP'P_{1})\Big]}{\vert \mathcal{A}(B_{b} \to B_{1}PP'P_{1})\vert^{2}+ \vert \mathcal{A}(\bar{B_{b}} \to \bar{B_{1}}\bar{P}\bar{P'}\bar{P_{1}})\vert^{2}},\label{eq:ACP}
\end{align}
where $\textbf{J}$ is 
the Jarlskog invariant given by $\text{Im}(V_{ub}V_{ud}^{*}V_{tb}^{*}V_{td})=
-\text{Im}(V_{ub}V_{us}^{*}V_{tb}^{*}V_{ts})=\textbf{J}$. Neglecting any $U$-spin breaking effects from phase space factors, we can relate the $CP$-asymmetry in $\Delta \vert S \vert =1$ and $\Delta S =0$ decay modes using Eq~\eqref{ratio rel 2}, 
\begin{align}
\label{ratio rel 2}
     \frac{A_{CP}(B_{bi} \to B_{1}^{a}P^{b}P^{'c}P_{1}^{d})\vert_{\Delta S}}
{A_{CP}(B_{bj} \to B_{1}^{e}P^{f}P^{'g}P_{1}^{h})\vert_{\Delta S'}}\!
\simeq-\frac{\tau_{\mathcal{B}_{bi}}}
{\tau_{\mathcal{B}_{bj}}}
\frac{Br(B_{bj} \to B_{1}^{e}P^{f}P^{'g}P_{1}^{h})} 
{Br(B_{bi} \to B_{1}^{a}P^{b}P^{'c}P_{1}^{d}))},  \qquad \Delta S\neq\Delta S'
\end{align}
 where $i$, $j$, and $a,\,e$  $b,\,f$, $c,\,g$, $d,\, h$ indices correspond to the various 
 baryon and mesons in the initial and final states and the negative sign accounts for the relative sign between the Jarlskog factor in $\Delta \vert S \vert =1$ and $\Delta S =0$ decay modes. These $A_{CP}$-relations can be read off from Table~\eqref{Tab:2} by using Eq~\eqref{ratio rel 2}. As reported by LHCb, the $\Lambda^{0}_{b}\to p^{+}\pi^{-}K^{-}\pi^{+}$ decay as well as other four-body bottom baryon decays occur primarily through hadronic resonances that individually decay into
two or three final-state particles. The contribution from multiple resonances results in substantial variation in associated strong phases and relative strengths of
tree ($\mathcal{A}_{T}$) and penguin amplitudes ($\mathcal{A}_{P}$) leading to differences in the $CP$-asymmetry across the final-state phase space of the bottom
baryon decay. Therefore, it is reasonable to investigate local $CP$-asymmetries near resonant regions in such decays, as will be explored in the following. If all the final-state mesons originate from a system with $U$-spin = 1/2,  as observed in one of the four phase-space regions of the decay $\Lambda^{0}_{b}\to p^{+}\pi^{-}K^{-}\pi^{+}$~\cite{LHCb:2025ray}, the $A^{\frac{3}{ 2},1,\pm}(\frac{1}{2})$ $U$-spin subamplitudes can be dropped. This simplification allows us to describe decays in Table~\eqref{Tab:2} in terms of 4 $U$-spin subamplitudes, two for each class of decays with a positively or negatively charged baryon in the final state. This leads to  additional $CP$-asymmetry relations that are given by, 
\begin{subequations}
\label{addn rel}
\begin{align}
    A_{CP}(\Lambda^{0}_{b}\to p^{+}(K^{-}K^{-}K^{+})) &\simeq A_{CP}(\Lambda^{0}_{b}\to p^{+}(\pi^{-}K^{-}\pi^{+})),\nonumber \\&
    = - A_{CP}(\Xi_{b}^{0}\to \Sigma^{+}(\pi^{-}\pi^{-}\pi^{+})) \frac{\tau_{\Lambda^{0}_{b}}}{\tau_{\Xi^{0}_{b}}}\frac{Br(\Xi_{b}^{0}\to \Sigma^{+}\pi^{-}\pi^{-}\pi^{+})}{Br(\Lambda^{0}_{b}\to p^{+}K^{-}K^{-}K^{+})}\nonumber\\&
    =- \frac{1}{2}A_{CP}(\Xi_{b}^{0}\to \Sigma^{+}(\pi^{-}K^{-}K^{+})) \frac{\tau_{\Lambda^{0}_{b}}}{\tau_{\Xi^{0}_{b}}}\frac{Br(\Xi_{b}^{0}\to \Sigma^{+}\pi^{-}K^{-}K^{+})}{Br(\Lambda^{0}_{b}\to p^{+}K^{-}K^{-}K^{+})}\label{addn rel 1}\\
      A_{CP}(\Lambda^{0}_{b}\to p^{+}(\pi^{-}\pi^{-}\pi^{+}))&\simeq A_{CP}(\Lambda^{0}_{b}\to p^{+}(\pi^{-}K^{-}K^{+}))
\nonumber\\&
 = -A_{CP}(\Xi_{b}^{0}\to \Sigma^{+}(K^{-}K^{-}K^{+}))\frac{\tau_{\Lambda^{0}_{b}}}{\tau_{\Xi^{0}_{b}}} \frac{Br(\Xi_{b}^{0}\to \Sigma^{+}K^{-}K^{-}K^{+})}{Br(\Lambda^{0}_{b}\to p^{+}\pi^{-}\pi^{-}\pi^{+})}\nonumber\\&
 =-\frac{1}{2}A_{CP}(\Xi_{b}^{0}\to \Sigma^{+}(\pi^{-}K^{-}\pi^{+}))\frac{\tau_{\Lambda^{0}_{b}}}{\tau_{\Xi^{0}_{b}}} \frac{Br(\Xi_{b}^{0}\to \Sigma^{+}\pi^{-}K^{-}\pi^{+})}{Br(\Lambda^{0}_{b}\to p^{+}\pi^{-}\pi^{-}\pi^{+})}\label{addn rel 2}\\
 A_{CP}(\Lambda^{0}_{b}\to \Sigma^{+}(\pi^{-}\pi^{-}\pi^{+}))&\simeq A_{CP}(\Lambda^{0}_{b}\to \Sigma^{+}(\pi^{-}K^{-}K^{+}))
\nonumber\\&
 = -A_{CP}(\Xi_{b}^{0}\to p^{+}(K^{-}K^{-}K^{+}))\frac{\tau_{\Lambda^{0}_{b}}}{\tau_{\Xi^{0}_{b}}} \frac{Br(\Xi_{b}^{0}\to p^{+}K^{-}K^{-}K^{+})}{Br(\Lambda^{0}_{b}\to \Sigma^{+}\pi^{-}\pi^{-}\pi^{+})}\nonumber\\&
 =-\frac{1}{2}A_{CP}(\Xi_{b}^{0}\to p^{+}(\pi^{-}K^{-}\pi^{+}))\frac{\tau_{\Lambda^{0}_{b}}}{\tau_{\Xi^{0}_{b}}} \frac{Br(\Xi_{b}^{0}\to p^{+}\pi^{-}K^{-}\pi^{+})}{Br(\Lambda^{0}_{b}\to \Sigma^{+}\pi^{-}\pi^{-}\pi^{+})}\label{addn rel 3}
 \end{align} 
\end{subequations}
Measurements of the following branching ratios of $\Lambda^{0}_{b}$-baryon and $\Xi^{0}_{b}$-baryon are available from LHCb~\cite{ParticleDataGroup:2024cfk},
\begin{align}
Br(\Lambda^{0}_{b}\to p^{+}K^{-}\pi^{-}\pi^{+})=(5.1 \pm 0.5)\times10^{-5} \nonumber\\
Br(\Lambda^{0}_{b}\to p^{+}K^{-}\pi^{-}K^{+})=(4.1\pm 0.6)\times 10^{-6} \nonumber\\ 
Br(\Lambda^{0}_{b}\to p^{+}K^{-}K^{-}K^{+})=(1.27\pm 0.13)\times 10^{-5}\\
Br(\Lambda^{0}_{b}\to p^{+}\pi^{-}\pi^{-}\pi^{+})=(2.12\pm 0.21)\times 10^{-5}\nonumber\\
Br(\Xi^{0}_{b}\to p^{+}K^{-}\pi^{-}\pi^{+})\times \frac{F(b\to \Xi_{b}^{0})}{F(b\to \Lambda^{0}_{b})}=(1.9 \pm 0.4)\times10^{-6} \nonumber\\
Br(\Xi^{0}_{b}\to p^{+}K^{-}K^{-}\pi^{+})\times \frac{F(b\to \Xi_{b}^{0})}{F(b\to \Lambda^{0}_{b})}=(1.73 \pm 0.31)\times10^{-6} \nonumber\\
Br(\Xi^{0}_{b}\to p^{+}K^{-}K^{-}K^{+})\times \frac{F(b\to \Xi_{b}^{0})}{F(b\to \Lambda^{0}_{b})}=(1.8 \pm 1.0)\times10^{-7} \nonumber
\end{align}
where $F(b\to B_{b})$ denotes the fragmentation fraction of $b$-quarks into a $B_{b}$-baryon. From the available local $CP$-asymmetry measurements in bottom baryon decays~\cite{LHCb:2019jyj,LHCb:2025ray}, we identify that $CP$-asymmetry in $\Lambda_{b}^{0}\to p^{+}K_{1}^{-}(1410)$ quasi-two body decay~\cite{LHCb:2019jyj}, 
\begin{align}
    A_{CP}(\Lambda_{b}^{0}\to p^{+} (\pi^{-}K^{-}\pi^{+}))=( 4.7 \pm 3.5 \pm 0.8) \%,
\end{align} can be translated into local $CP$-asymmetry in $\Lambda_{b}^{0}\to p^{+}\pi^{-}K^{-}\pi^{+}$ transition since $K_{1}^{-}(1410)$ decays strongly to $\pi^{+}K^{-}\pi^{-}$~\cite{ParticleDataGroup:2024cfk}. Moreover, $K_{1}^{-}(1410)\to K^{+}K^{-}K^{-}$ strong decay is also possible~\cite{LHCb:2021uow} and therefore based on Eq~\eqref{addn rel 1} we expect that local $CP$-asymmetry in $\Lambda^{0}_{b}\to p^{+}K^{-}K^{-}K^{+}$ transition is,
\begin{align}
A_{CP}(\Lambda_{b}^{0}\to p^{+} (K^{-}K^{-}K^{+})_{K_{1}})= A_{CP}(\Lambda_{b}^{0}\to p^{+} (\pi^{-}K^{-}\pi^{+})_{K_{1}})\simeq (4.7 \pm 3.6)\%
\end{align}  
We observe that the predicted value of $A_{CP}(\Lambda_{b}^{0}\to p^{+} (\pi^{-}K^{-}\pi^{+})_{K_{1}})$ has the same sign as the recently measured local $A_{CP}$~\cite{LHCb:2025ray}
\begin{align}
A_{CP}(\Lambda_{b}^{0}\to p^{+} (\pi^{-}K^{-}\pi^{+}))\vert_{m_{\pi^{+}K^{-}\pi^{-}}<\, 2.0 \, \text{GeV}}=(2.0\pm 1.2 \pm 0.3)\%,
\end{align}
although the central values differ by roughly a factor of two.
Similarly, we focus on $CP$-violation in the quasi-two body $\Lambda^{0}_{b}\to p^{+} a_{1}^{-}(1260)$ decay where we note that $a_{1}^{-}(1260)$ subsequently can decay strongly to either $\pi^{+}\pi^{-}\pi^{-}$ or $K^{+}\pi^{-}K^{-}$~\cite{ParticleDataGroup:2024cfk}. We use the measured $CP$-asymmetry in $\Lambda^{0}_{b}\to p^{+}a_{1}^{-}(1260)$-decay~\cite{LHCb:2019jyj},
\begin{align}
A_{CP}(\Lambda^{0}_{b}\to p^{+} a_{1}^{-}(1260))=(-1.5 \pm 4.2 \pm 0.6)\%
\end{align}
and with the help of Eq~\eqref{addn rel 2} also predict local $CP$-asymmetry in $\Lambda^{0}_{b}\to p^{+}\pi^{-}K^{-}K^{+}$ and $\Lambda^{0}_{b}\to p^{+}\pi^{-}\pi^{-}\pi^{+}$,
\begin{align}
    A_{CP}(\Lambda^{0}_{b}\to p^{+}(\pi^{-}\pi^{-}\pi^{+})_{a_{1}})=A_{CP}(\Lambda^{0}_{b}\to p^{+}(\pi^{-}K^{-}K^{+})_{a_{1}})\simeq(-1.5\pm 4.2)\%.
\end{align}
Once the branching fractions and local $CP$ asymmetries of additional $\Lambda^{0}_{b}$ and $\Xi^{0}_{b}$ decay modes, such as $\Lambda^{0}_{b} \to \Sigma^{+} \pi^{-} \pi^{-} \pi^{+}$ and $\Lambda^{0}_{b} \to \Sigma^{+} \pi^{-} K^{-} K^{+}$ become available, they can be used to express the local-$CP$ asymmetries in related $\Xi^{0}_{b}$ decay modes: $\Xi^{0}_{b} \to p^{+} K^{-} K^{-} K^{+}$ and $\Xi^{0}_{b} \to p^{+} \pi^{-} K^{-} \pi^{+}$ with the help of Eq~\eqref{addn rel 3}. The $CP$-asymmetry relations inferred from Table~\eqref{Tab:2} can also be tested by experiments in near future. We now discuss the effect of first order breaking of $U$-spin originating form unequal masses of the $d$ an $s$ quark. $U$-spin breaking is introduced by a triplet under $U$-spin ($U_{\text{brk}}=1,U^{3}_{\text{brk}}=0$) and by taking the tensor product with the unbroken Hamiltonian that transforms as a doublet. The effect of this $U$-spin breaking is a correction to the unbroken Hamiltonian given by,
\begin{align}
    \mathcal{H}_{\text{brk}}^{\Delta \vert S \vert=0,1}\propto \mathcal{\epsilon}(1,0)\otimes(\frac{1}{2},\pm \frac{1}{2})=\epsilon\Big(\sqrt{\frac{2}{3}}(\frac{3}{2},\pm \frac{1}{2})\mp \sqrt{\frac{1}{3}}(\frac{1}{2},\pm \frac{1}{2})\Big)
\end{align}
This induces non-zero dynamical coefficients for the $U$-spin subamplitudes $ A^{\frac{1}{ 2},1,\pm}(\frac{3}{2}), \, A^{\frac{3}{ 2},1,\pm}(\frac{3}{2}),$ $ A^{\frac{3}{ 2},2,\pm}(\frac{3}{2})$ while  $A^{\frac{1}{ 2},0}(\frac{1}{2}),\, 
     A^{\frac{1}{ 2},1}(\frac{1}{2}),\, 
     A^{\frac{3}{ 2},1}(\frac{1}{2})$ pick up $U$-spin breaking corrections. Collectively, we can parametrize these $U$-spin breaking amplitudes as, 
  \begin{align}
      A_{\text{brk}}^{\frac{1}{ 2},0,\pm}(\frac{1}{2})=\frac{\epsilon_{1}^{\pm}}{\sqrt{3}} A^{\frac{1}{ 2},0,\pm}(\frac{1}{2}),\qquad
     A_{\text{brk}}^{\frac{1}{ 2},1,\pm}(\frac{1}{2})= \frac{\epsilon_{2}^{\pm}}{\sqrt{3}}A^{\frac{1}{ 2},1,\pm}(\frac{1}{2})\nonumber \\
      A_{\text{brk}}^{\frac{1}{ 2},1,\pm}(\frac{3}{2})=\frac{\sqrt{2}\epsilon_{3}^{\pm}}{\sqrt{3}}A^{\frac{1}{ 2},1,\pm}(\frac{3}{2}),\qquad
       A_{\text{brk}}^{\frac{3}{ 2},1,\pm}(\frac{1}{2})=\frac{\epsilon_{4}^{\pm}}{\sqrt{3}}A^{\frac{3}{ 2},1,\pm}(\frac{1}{2}),\\
        A_{\text{brk}}^{\frac{3}{ 2},1,\pm}(\frac{3}{2})=\frac{\sqrt{2}\epsilon_{5}^{\pm}}{\sqrt{3}}A^{\frac{3}{ 2},1,\pm}(\frac{3}{2}),\qquad
         A_{\text{brk}}^{\frac{3}{ 2},2,\pm}(\frac{3}{2})=\frac{\sqrt{2}\epsilon_{6}^{\pm}}{\sqrt{3}}A^{\frac{3}{ 2},2,\pm}(\frac{3}{2})\nonumber          
  \end{align}   
The combined result of leading order unbroken Hamiltonian and first order $U$-spin breaking contribution to the decay amplitudes is given in Appendix~\eqref{App:2}. The first order $U$-spin breaking corrections have the same structures but opposite signs for the decay modes related by $U$-spin as found in~\cite{Gronau:2013mda}. This increases the rank of the coefficient matrices to 12 and the leading order relations between $U$-spin conjugate amplitude pairs no longer hold. Naturally, the leading order $CP$-asymmetry relations evident from Table~\eqref{Tab:2} and Eq~\eqref{addn rel} also break down once $U$-spin breaking effects are considered.  Any reduction among the set of $U$-spin subamplitudes would require knowledge of their relative magnitudes, which lies beyond the scope of the present work.
 \section{Conclusion}
 We have investigated charmless bottom baryon decays into fully charged four-body final states within the realm of $U$-spin symmetry. Bearing in mind that significant $U$-spin breaking is observed in $B$-meson decays, we set up a framework to analyze bottom baryon decays where $U$-spin symmetry can be arbitrarily broken. We obtained amplitude relations between decay modes and based on those relations predicted $CP$ asymmetry relations between a number of decay modes. By making the additional assumption that all final-state mesons emerge from a $U$-spin doublet, we demonstrated that this extended framework allows for numerical predictions of localized $CP$ asymmetries in specific regions of phase space, which hold to a good accuracy in practice. The framework also predicts the relative signs of the $CP$ asymmetries, an important feature that can be obscured in global $CP$ asymmetry measurements due to averaging over the entire phase space of the decay. While $U$-spin breaking effect spoils the relations between observables, the predictions offer a roadmap for probing $CP$ violation in bottom baryon decays and provide a systematic approach to test the validity of $U$-spin symmetry in future experimental measurements. 
 \vspace{-1cm}
\begin{appendix}
\section{}
\label{App:1}
\resizebox{0.9\linewidth}{!}{
  \begin{minipage}{\linewidth}
$U$-spin amplitude decomposition of $B_{b} \to B_{1}^{+}PP'P_{1}$ decay modes for general Hamiltonian: 
\begin{align}
\begin{pmatrix}
\mathcal{A}(\Lambda^{0}_{b}\to p^{+}K^{-}\pi^{-}\pi^{+}) \\
\mathcal{A}(\Lambda^{0}_{b}\to \Sigma^{+}\pi^{-}\pi^{-}\pi^{+})\\
\mathcal{A}(\Xi_{b}^{0}\to \Sigma^{+}K^{-}K^{-}K^{+})\\
\mathcal{A}(\Lambda_{b}^{0}\to p^{+}K^{-}K^{-}K^{+})\\
\mathcal{A}(\Lambda_{b}^{0}\to \Sigma^{+}K^{-}\pi^{-}K^{+})\\
\mathcal{A}(\Xi_{b}^{0}\to p^{+}K^{-}K^{-}\pi^{+})\\
\mathcal{A}(\Xi_{b}^{0}\to \Sigma^{+}\pi^{-}K^{-}\pi^{+})\\
\mathcal{A}(\Xi^{0}_{b}\to \Sigma^{+}K^{-}\pi^{-}K^{+}) \\
\mathcal{A}(\Xi^{0}_{b}\to p^{+}K^{-}K^{-}K^{+})\\
\mathcal{A}(\Lambda_{b}^{0}\to p^{+}\pi^{-}\pi^{-}\pi^{+})\\
\mathcal{A}(\Xi_{b}^{0}\to \Sigma^{+}\pi^{-}\pi^{-}\pi^{+})\\
\mathcal{A}(\Xi_{b}^{0}\to p^{+}K^{-}\pi^{-}\pi^{+})\\
\mathcal{A}(\Lambda_{b}^{0}\to \Sigma^{+}\pi^{-}\pi^{-}K^{+})\\
\mathcal{A}(\Lambda_{b}^{0}\to p^{+}\pi^{-}K^{-}K^{+})\\
\end{pmatrix}
=\left(
\begin{array}{ccccccc}
 \frac{1}{\sqrt{6}} & -\frac{1}{3 \sqrt{2}} & -\frac{1}{3} & \frac{1}{3} & \frac{\sqrt{2}}{3} & -\sqrt{\frac{2}{15}} & -\frac{1}{\sqrt{5}} 
   \\
 -\frac{1}{\sqrt{3}} & -\frac{1}{3} & -\frac{\sqrt{2}}{3} & -\frac{1}{3 \sqrt{2}} & -\frac{1}{3} & -\frac{1}{\sqrt{15}} & -\frac{1}{\sqrt{10}}  \\
 0 & \frac{2}{3} & -\frac{\sqrt{2}}{3} & -\frac{1}{3 \sqrt{2}} & \frac{1}{6} & -\frac{\sqrt{\frac{3}{5}}}{2} & \frac{1}{\sqrt{10}}  \\
 \frac{1}{\sqrt{3}} & -\frac{1}{3} & -\frac{\sqrt{2}}{3} & -\frac{1}{3 \sqrt{2}} & -\frac{1}{3} & \frac{1}{\sqrt{15}} & \frac{1}{\sqrt{10}} 
   \\
 -\frac{1}{\sqrt{6}} & -\frac{1}{3 \sqrt{2}} & -\frac{1}{3} & \frac{1}{3} & \frac{\sqrt{2}}{3} & \sqrt{\frac{2}{15}} & \frac{1}{\sqrt{5}} 
   \\
 0 & 0 & 0 & -\frac{1}{\sqrt{2}} & \frac{1}{2} & \frac{\sqrt{\frac{3}{5}}}{2} & -\frac{1}{\sqrt{10}} \\
 0 & \frac{\sqrt{2}}{3} & -\frac{1}{3} & \frac{1}{3} & -\frac{1}{3 \sqrt{2}} & \sqrt{\frac{3}{10}} & -\frac{1}{\sqrt{5}} \\
 -\frac{1}{\sqrt{6}} & \frac{1}{3 \sqrt{2}} & -\frac{1}{3} & -\frac{1}{3} & \frac{\sqrt{2}}{3} & -\sqrt{\frac{2}{15}} & \frac{1}{\sqrt{5}}
   \\
 \frac{1}{\sqrt{3}} & \frac{1}{3} & -\frac{\sqrt{2}}{3} & \frac{1}{3 \sqrt{2}} & -\frac{1}{3} & -\frac{1}{\sqrt{15}} & \frac{1}{\sqrt{10}} 
   \\
 0 & -\frac{2}{3} & -\frac{\sqrt{2}}{3} & \frac{1}{3 \sqrt{2}} & \frac{1}{6} & -\frac{\sqrt{\frac{3}{5}}}{2} & -\frac{1}{\sqrt{10}}  \\
 -\frac{1}{\sqrt{3}} & \frac{1}{3} & -\frac{\sqrt{2}}{3} & \frac{1}{3 \sqrt{2}} & -\frac{1}{3} & \frac{1}{\sqrt{15}} & -\frac{1}{\sqrt{10}} 
   \\
 \frac{1}{\sqrt{6}} & \frac{1}{3 \sqrt{2}} & -\frac{1}{3} & -\frac{1}{3} & \frac{\sqrt{2}}{3} & \sqrt{\frac{2}{15}} & -\frac{1}{\sqrt{5}}
   \\
 0 & 0 & 0 & \frac{1}{\sqrt{2}} & \frac{1}{2} & \frac{\sqrt{\frac{3}{5}}}{2} & \frac{1}{\sqrt{10}}  \\
 0 & -\frac{\sqrt{2}}{3} & -\frac{1}{3} & -\frac{1}{3} & -\frac{1}{3 \sqrt{2}} & \sqrt{\frac{3}{10}} & \frac{1}{\sqrt{5}} \\
\end{array}
\right)
\begin{pmatrix}
    A^{\frac{1}{ 2},0,+}(\frac{1}{2})\\
     A^{\frac{1}{ 2},1,+}(\frac{1}{2})\\
      A^{\frac{1}{ 2},1,+}(\frac{3}{2})\\
       A^{\frac{3}{ 2},1,+}(\frac{1}{2})\\
        A^{\frac{3}{ 2},1,+}(\frac{3}{2})\\
         A^{\frac{3}{ 2},2,+}(\frac{3}{2})\\
           A^{\frac{3}{ 2},2,+}(\frac{5}{2})
\end{pmatrix}
\end{align}
$U$-spin amplitude decomposition of $B_{b} \to B_{1}^{-}PP'P_{1}$ decay modes for general Hamiltonian:
\begin{align}
\begin{pmatrix}
\mathcal{A}(\Lambda^{0}_{b}\to \Xi^{-}K^{+}K^{+}K^{-})\\
\mathcal{A}(\Lambda^{0}_{b}\to \Sigma^{-}K^{+}\pi^{+}K^{-})\\
\mathcal{A}(\Xi_{b}^{0}\to \Xi^{-}\pi^{+}\pi^{+}\pi^{-})\\
\mathcal{A}(\Lambda_{b}^{0}\to \Xi^{-}K^{+}\pi^{+}\pi^{-})\\
\mathcal{A}(\Lambda_{b}^{0}\to \Sigma^{-}\pi^{+}\pi^{+}\pi^{-})\\
\mathcal{A}(\Xi_{b}^{0}\to \Xi^{-}\pi^{+}K^{+}K^{-})\\
\mathcal{A}(\Xi_{b}^{0}\to \Sigma^{-}\pi^{+}\pi^{+}K^{-})\\
\mathcal{A}(\Xi^{0}_{b}\to \Sigma^{-}\pi^{+}\pi^{+}\pi^{-})\\
\mathcal{A}(\Xi^{0}_{b}\to \Xi^{-}\pi^{+}K^{+}\pi^{-})\\
\mathcal{A}(\Lambda_{b}^{0}\to \Sigma^{-}K^{+}K^{+}K^{-})\\
\mathcal{A}(\Xi_{b}^{0}\to \Sigma^{-}\pi^{+}K^{+}K^{-})\\
\mathcal{A}(\Xi_{b}^{0}\to \Xi^{-}K^{+}K^{+}K^{-})\\
\mathcal{A}(\Lambda_{b}^{0}\to \Sigma^{-}\pi^{+}K^{+}\pi^{-})\\
\mathcal{A}(\Lambda_{b}^{0}\to \Xi^{-}K^{+}K^{+}\pi^{-})\\
\end{pmatrix}
=\left(
\begin{array}{ccccccc}
  -\frac{1}{\sqrt{3}} & -\frac{1}{3} & -\frac{\sqrt{2}}{3} & -\frac{1}{3 \sqrt{2}} & -\frac{1}{3} & -\frac{1}{\sqrt{15}} &
   -\frac{1}{\sqrt{10}} \\
  \frac{1}{\sqrt{6}} & -\frac{1}{3 \sqrt{2}} & -\frac{1}{3} & \frac{1}{3} & \frac{\sqrt{2}}{3} & -\sqrt{\frac{2}{15}} & -\frac{1}{\sqrt{5}}
   \\
  0 & \frac{2}{3} & -\frac{\sqrt{2}}{3} & -\frac{1}{3 \sqrt{2}} & \frac{1}{6} & -\frac{\sqrt{\frac{3}{5}}}{2} & \frac{1}{\sqrt{10}} \\
  -\frac{1}{\sqrt{6}} & -\frac{1}{3 \sqrt{2}} & -\frac{1}{3} & \frac{1}{3} & \frac{\sqrt{2}}{3} & \sqrt{\frac{2}{15}} & \frac{1}{\sqrt{5}}
   \\
  \frac{1}{\sqrt{3}} & -\frac{1}{3} & -\frac{\sqrt{2}}{3} & -\frac{1}{3 \sqrt{2}} & -\frac{1}{3} & \frac{1}{\sqrt{15}} & \frac{1}{\sqrt{10}}
   \\
  0 & \frac{\sqrt{2}}{3} & -\frac{1}{3} & \frac{1}{3} & -\frac{1}{3 \sqrt{2}} & \sqrt{\frac{3}{10}} & -\frac{1}{\sqrt{5}} \\
  0 & 0 & 0 & -\frac{1}{\sqrt{2}} & \frac{1}{2} & \frac{\sqrt{\frac{3}{5}}}{2} & -\frac{1}{\sqrt{10}} \\
  \frac{1}{\sqrt{3}} & \frac{1}{3} & -\frac{\sqrt{2}}{3} & \frac{1}{3 \sqrt{2}} & -\frac{1}{3} & -\frac{1}{\sqrt{15}} & \frac{1}{\sqrt{10}}
   \\
  -\frac{1}{\sqrt{6}} & \frac{1}{3 \sqrt{2}} & -\frac{1}{3} & -\frac{1}{3} & \frac{\sqrt{2}}{3} & -\sqrt{\frac{2}{15}} & \frac{1}{\sqrt{5}}
   \\
  0 & -\frac{2}{3} & -\frac{\sqrt{2}}{3} & \frac{1}{3 \sqrt{2}} & \frac{1}{6} & -\frac{\sqrt{\frac{3}{5}}}{2} & -\frac{1}{\sqrt{10}} \\
  \frac{1}{\sqrt{6}} & \frac{1}{3 \sqrt{2}} & -\frac{1}{3} & -\frac{1}{3} & \frac{\sqrt{2}}{3} & \sqrt{\frac{2}{15}} & -\frac{1}{\sqrt{5}}
   \\
  -\frac{1}{\sqrt{3}} & \frac{1}{3} & -\frac{\sqrt{2}}{3} & \frac{1}{3 \sqrt{2}} & -\frac{1}{3} & \frac{1}{\sqrt{15}} & -\frac{1}{\sqrt{10}}
   \\
  0 & -\frac{\sqrt{2}}{3} & -\frac{1}{3} & -\frac{1}{3} & -\frac{1}{3 \sqrt{2}} & \sqrt{\frac{3}{10}} & \frac{1}{\sqrt{5}} \\
  0 & 0 & 0 & \frac{1}{\sqrt{2}} & \frac{1}{2} & \frac{\sqrt{\frac{3}{5}}}{2} & \frac{1}{\sqrt{10}} \\
\end{array}
\right)
\begin{pmatrix}
    A^{\frac{1}{ 2},0,-}(\frac{1}{2})\\
     A^{\frac{1}{ 2},1,-}(\frac{1}{2})\\
      A^{\frac{1}{ 2},1,-}(\frac{3}{2})\\
       A^{\frac{3}{ 2},1,-}(\frac{1}{2})\\
        A^{\frac{3}{ 2},1,-}(\frac{3}{2})\\
         A^{\frac{3}{ 2},2,-}(\frac{3}{2})\\
           A^{\frac{3}{ 2},2,-}(\frac{5}{2})
\end{pmatrix}
\end{align}
\end{minipage}
}
\newpage
$U$-spin amplitude decomposition of $B_{b} \to B_{1}PP'P_{1}$ decay modes using the SM Hamiltonian,
\begin{align}
    \begin{pmatrix}
      \mathcal{A}(\Lambda^{0}_{b}\to p^{+}K^{-}\pi^{-}\pi^{+} )\\
\mathcal{A}(\Lambda^{0}_{b}\to \Xi^{-}K^{+}K^{+}K^{-})\\
\mathcal{A}(\Lambda^{0}_{b}\to \Sigma^{+}\pi^{-}\pi^{-}\pi^{+})\\
\mathcal{A}(\Lambda^{0}_{b}\to \Sigma^{-}K^{+}\pi^{+}K^{-})\\
\mathcal{A}(\Xi_{b}^{0}\to \Xi^{-}\pi^{+}\pi^{+}\pi^{-})\\
\mathcal{A}(\Xi_{b}^{0}\to \Sigma^{+}K^{-}K^{-}K^{+})\\
\mathcal{A}(\Lambda_{b}^{0}\to p^{+}K^{-}K^{-}K^{+})\\
\mathcal{A}(\Lambda_{b}^{0}\to \Xi^{-}K^{+}\pi^{+}\pi^{-})\\
\mathcal{A}(\Lambda_{b}^{0}\to \Sigma^{+}K^{-}\pi^{-}K^{+})\\
\mathcal{A}(\Lambda_{b}^{0}\to \Sigma^{-}\pi^{+}\pi^{+}\pi^{-})\\
\mathcal{A}(\Xi_{b}^{0}\to p^{+}K^{-}K^{-}\pi^{+})\\
\mathcal{A}(\Xi_{b}^{0}\to \Xi^{-}\pi^{+}K^{+}K^{-})\\
\mathcal{A}(\Xi_{b}^{0}\to \Sigma^{+}\pi^{-}K^{-}\pi^{+})\\
\mathcal{A}(\Xi_{b}^{0}\to \Sigma^{-}\pi^{+}\pi^{+}K^{-})\\ 
\mathcal{A}(\Xi^{0}_{b}\to \Sigma^{+}K^{-}\pi^{-}K^{+} )\\
\mathcal{A}(\Xi^{0}_{b}\to \Sigma^{-}\pi^{+}\pi^{+}\pi^{-})\\
\mathcal{A}(\Xi^{0}_{b}\to p^{+}K^{-}K^{-}K^{+})\\
\mathcal{A}(\Xi^{0}_{b}\to \Xi^{-}\pi^{+}K^{+}\pi^{-})\\
\mathcal{A}(\Lambda_{b}^{0}\to \Sigma^{-}K^{+}K^{+}K^{-})\\
\mathcal{A}(\Lambda_{b}^{0}\to p^{+}\pi^{-}\pi^{-}\pi^{+})\\
\mathcal{A}(\Xi_{b}^{0}\to \Sigma^{+}\pi^{-}\pi^{-}\pi^{+})\\
\mathcal{A}(\Xi_{b}^{0}\to \Sigma^{-}\pi^{+}K^{+}K^{-})\\
\mathcal{A}(\Xi_{b}^{0}\to p^{+}K^{-}\pi^{-}\pi^{+})\\
\mathcal{A}(\Xi_{b}^{0}\to \Xi^{-}K^{+}K^{+}K^{-})\\
\mathcal{A}(\Lambda_{b}^{0}\to \Sigma^{+}\pi^{-}\pi^{-}K^{+})\\
\mathcal{A}(\Lambda_{b}^{0}\to \Sigma^{-}\pi^{+}K^{+}\pi^{-})\\
\mathcal{A}(\Lambda_{b}^{0}\to p^{+}\pi^{-}K^{-}K^{+})\\
\mathcal{A}(\Lambda_{b}^{0}\to \Xi^{-}K^{+}K^{+}\pi^{-})\\
    \end{pmatrix}
=\mathcal{C}_{T (P) }
\left(
\begin{array}{cccccc}
 \frac{1}{\sqrt{6}} & -\frac{1}{3 \sqrt{2}} & \frac{1}{3} & 0 & 0 & 0 \\
 0 & 0 & 0 & -\frac{1}{\sqrt{3}} & -\frac{1}{3} & -\frac{1}{3 \sqrt{2}} \\
 -\frac{1}{\sqrt{3}} & -\frac{1}{3} & -\frac{1}{3 \sqrt{2}} & 0 & 0 & 0 \\
 0 & 0 & 0 & \frac{1}{\sqrt{6}} & -\frac{1}{3 \sqrt{2}} & \frac{1}{3} \\
 0 & 0 & 0 & 0 & \frac{2}{3} & -\frac{1}{3 \sqrt{2}} \\
 0 & \frac{2}{3} & -\frac{1}{3 \sqrt{2}} & 0 & 0 & 0 \\
 \frac{1}{\sqrt{3}} & -\frac{1}{3} & -\frac{1}{3 \sqrt{2}} & 0 & 0 & 0 \\
 0 & 0 & 0 & -\frac{1}{\sqrt{6}} & -\frac{1}{3 \sqrt{2}} & \frac{1}{3} \\
 -\frac{1}{\sqrt{6}} & -\frac{1}{3 \sqrt{2}} & \frac{1}{3} & 0 & 0 & 0 \\
 0 & 0 & 0 & \frac{1}{\sqrt{3}} & -\frac{1}{3} & -\frac{1}{3 \sqrt{2}} \\
 0 & 0 & -\frac{1}{\sqrt{2}} & 0 & 0 & 0 \\
 0 & 0 & 0 & 0 & \frac{\sqrt{2}}{3} & \frac{1}{3} \\
 0 & \frac{\sqrt{2}}{3} & \frac{1}{3} & 0 & 0 & 0 \\
 0 & 0 & 0 & 0 & 0 & -\frac{1}{\sqrt{2}} \\
 -\frac{1}{\sqrt{6}} & \frac{1}{3 \sqrt{2}} & -\frac{1}{3} & 0 & 0 & 0 \\
 0 & 0 & 0 & \frac{1}{\sqrt{3}} & \frac{1}{3} & \frac{1}{3 \sqrt{2}} \\
 \frac{1}{\sqrt{3}} & \frac{1}{3} & \frac{1}{3 \sqrt{2}} & 0 & 0 & 0 \\
 0 & 0 & 0 & -\frac{1}{\sqrt{6}} & \frac{1}{3 \sqrt{2}} & -\frac{1}{3} \\
 0 & 0 & 0 & 0 & -\frac{2}{3} & \frac{1}{3 \sqrt{2}} \\
 0 & -\frac{2}{3} & \frac{1}{3 \sqrt{2}} & 0 & 0 & 0 \\
 -\frac{1}{\sqrt{3}} & \frac{1}{3} & \frac{1}{3 \sqrt{2}} & 0 & 0 & 0 \\
 0 & 0 & 0 & \frac{1}{\sqrt{6}} & \frac{1}{3 \sqrt{2}} & -\frac{1}{3} \\
 \frac{1}{\sqrt{6}} & \frac{1}{3 \sqrt{2}} & -\frac{1}{3} & 0 & 0 & 0 \\
 0 & 0 & 0 & -\frac{1}{\sqrt{3}} & \frac{1}{3} & \frac{1}{3 \sqrt{2}} \\
 0 & 0 & \frac{1}{\sqrt{2}} & 0 & 0 & 0 \\
 0 & 0 & 0 & 0 & -\frac{\sqrt{2}}{3} & -\frac{1}{3} \\
 0 & -\frac{\sqrt{2}}{3} & -\frac{1}{3} & 0 & 0 & 0 \\
 0 & 0 & 0 & 0 & 0 & \frac{1}{\sqrt{2}} \\
\end{array}
\right)
\begin{pmatrix}
     A^{\frac{1}{ 2},0,+}(\frac{1}{2})\\
     A^{\frac{1}{ 2},1,+}(\frac{1}{2})\\
     A^{\frac{3}{ 2},1,+}(\frac{1}{2})\\
      A^{\frac{1}{ 2},0,-}(\frac{1}{2})\\
     A^{\frac{1}{ 2},1,-}(\frac{1}{2})\\
     A^{\frac{3}{ 2},1,-}(\frac{1}{2})\\
    \end{pmatrix}
\end{align}
where $\mathcal{C}_{T}$ and $\mathcal{C}_{P}$ are the Wilson Coefficients of tree operators ($O_{1}, \, O_{2}$) and penguin operators ($O_{3}\dots O_{10}$) respectively.  
\section{}
\label{App:2}
$U$-spin amplitude decomposition of $B_{b} \to B_{1}PP'P_{1}$ decay modes incorporating first order $U$-spin breaking to the SM Hamiltonian,
\begin{align}
\rotatebox{90}{
$
\resizebox{0.8\vsize}{!}{%
$
 \begin{pmatrix}
      \mathcal{A}(\Lambda^{0}_{b}\to p^{+}K^{-}\pi^{-}\pi^{+} \\
\mathcal{A}(\Lambda^{0}_{b}\to \Xi^{-}K^{+}K^{+}K^{-}\\
\mathcal{A}(\Lambda^{0}_{b}\to \Sigma^{+}\pi^{-}\pi^{-}\pi^{+})\\
\mathcal{A}(\Lambda^{0}_{b}\to \Sigma^{-}K^{+}\pi^{+}K^{-})\\
\mathcal{A}(\Xi_{b}^{0}\to \Xi^{-}\pi^{+}\pi^{+}\pi^{-})\\
\mathcal{A}(\Xi_{b}^{0}\to \Sigma^{+}K^{-}K^{-}K^{+})\\
\mathcal{A}(\Lambda_{b}^{0}\to p^{+}K^{-}K^{-}K^{+})\\
\mathcal{A}(\Lambda_{b}^{0}\to \Xi^{-}K^{+}\pi^{+}\pi^{-})\\
\mathcal{A}(\Lambda_{b}^{0}\to \Sigma^{+}K^{-}\pi^{-}K^{+})\\
\mathcal{A}(\Lambda_{b}^{0}\to \Sigma^{-}\pi^{+}\pi^{+}\pi^{-})\\
\mathcal{A}(\Xi_{b}^{0}\to p^{+}K^{-}K^{-}\pi^{+})\\
\mathcal{A}(\Xi_{b}^{0}\to \Xi^{-}\pi^{+}K^{+}K^{-})\\
\mathcal{A}(\Xi_{b}^{0}\to \Sigma^{+}\pi^{-}K^{-}\pi^{+})\\
\mathcal{A}(\Xi_{b}^{0}\to \Sigma^{-}\pi^{+}\pi^{+}K^{-})\\ 
\mathcal{A}(\Xi^{0}_{b}\to \Sigma^{+}K^{-}\pi^{-}K^{+} \\
\mathcal{A}(\Xi^{0}_{b}\to \Sigma^{-}\pi^{+}\pi^{+}\pi^{-}\\
\mathcal{A}(\Xi^{0}_{b}\to p^{+}K^{-}K^{-}K^{+})\\
\mathcal{A}(\Xi^{0}_{b}\to \Xi^{-}\pi^{+}K^{+}\pi^{-})\\
\mathcal{A}(\Lambda_{b}^{0}\to \Sigma^{-}K^{+}K^{+}K^{-})\\
\mathcal{A}(\Lambda_{b}^{0}\to p^{+}\pi^{-}\pi^{-}\pi^{+})\\
\mathcal{A}(\Xi_{b}^{0}\to \Sigma^{+}\pi^{-}\pi^{-}\pi^{+})\\
\mathcal{A}(\Xi_{b}^{0}\to \Sigma^{-}\pi^{+}K^{+}K^{-})\\
\mathcal{A}(\Xi_{b}^{0}\to p^{+}K^{-}\pi^{-}\pi^{+})\\
\mathcal{A}(\Xi_{b}^{0}\to \Xi^{-}K^{+}K^{+}K^{-})\\
\mathcal{A}(\Lambda_{b}^{0}\to \Sigma^{+}\pi^{-}\pi^{-}K^{+})\\
\mathcal{A}(\Lambda_{b}^{0}\to \Sigma^{-}\pi^{+}K^{+}\pi^{-})\\
\mathcal{A}(\Lambda_{b}^{0}\to p^{+}\pi^{-}K^{-}K^{+})\\
\mathcal{A}(\Lambda_{b}^{0}\to \Xi^{-}K^{+}K^{+}\pi^{-})\\
    \end{pmatrix}=
\left(
\begin{array}{cccccccccccc}
 \frac{\left(\epsilon _1\right){}^+}{3 \sqrt{2}}+\frac{1}{\sqrt{6}} & -\frac{\left(\epsilon _2\right){}^+}{3 \sqrt{6}}-\frac{1}{3 \sqrt{2}} & -\frac{1}{3}
   \sqrt{\frac{2}{3}} \left(\epsilon _3\right){}^+ & \frac{\left(\epsilon _4\right){}^+}{3 \sqrt{3}}+\frac{1}{3} & \frac{2 \left(\epsilon _5\right){}^+}{3 \sqrt{3}} &
   -\frac{2 \left(\epsilon _6\right){}^+}{3 \sqrt{5}} & 0 & 0 & 0 & 0 & 0 & 0 \\
 0 & 0 & 0 & 0 & 0 & 0 & -\frac{1}{3} \left(\epsilon _1\right){}^--\frac{1}{\sqrt{3}} & -\frac{\left(\epsilon _2\right){}^-}{3 \sqrt{3}}-\frac{1}{3} & -\frac{2
   \left(\epsilon _3\right){}^-}{3 \sqrt{3}} & -\frac{\left(\epsilon _4\right){}^-}{3 \sqrt{6}}-\frac{1}{3 \sqrt{2}} & -\frac{1}{3} \sqrt{\frac{2}{3}} \left(\epsilon
   _5\right){}^- & -\frac{1}{3} \sqrt{\frac{2}{5}} \left(\epsilon _6\right){}^- \\
 -\frac{1}{3} \left(\epsilon _1\right){}^+-\frac{1}{\sqrt{3}} & -\frac{\left(\epsilon _2\right){}^+}{3 \sqrt{3}}-\frac{1}{3} & -\frac{2 \left(\epsilon _3\right){}^+}{3
   \sqrt{3}} & -\frac{\left(\epsilon _4\right){}^+}{3 \sqrt{6}}-\frac{1}{3 \sqrt{2}} & -\frac{1}{3} \sqrt{\frac{2}{3}} \left(\epsilon _5\right){}^+ & -\frac{1}{3}
   \sqrt{\frac{2}{5}} \left(\epsilon _6\right){}^+ & 0 & 0 & 0 & 0 & 0 & 0 \\
 0 & 0 & 0 & 0 & 0 & 0 & \frac{\left(\epsilon _1\right){}^-}{3 \sqrt{2}}+\frac{1}{\sqrt{6}} & -\frac{\left(\epsilon _2\right){}^-}{3 \sqrt{6}}-\frac{1}{3 \sqrt{2}} &
   -\frac{1}{3} \sqrt{\frac{2}{3}} \left(\epsilon _3\right){}^- & \frac{\left(\epsilon _4\right){}^-}{3 \sqrt{3}}+\frac{1}{3} & \frac{2 \left(\epsilon _5\right){}^-}{3
   \sqrt{3}} & -\frac{2 \left(\epsilon _6\right){}^-}{3 \sqrt{5}} \\
 0 & 0 & 0 & 0 & 0 & 0 & 0 & \frac{2 \left(\epsilon _2\right){}^-}{3 \sqrt{3}}+\frac{2}{3} & -\frac{2 \left(\epsilon _3\right){}^-}{3 \sqrt{3}} & -\frac{\left(\epsilon
   _4\right){}^-}{3 \sqrt{6}}-\frac{1}{3 \sqrt{2}} & \frac{\left(\epsilon _5\right){}^-}{3 \sqrt{6}} & -\frac{\left(\epsilon _6\right){}^-}{\sqrt{10}} \\
 0 & \frac{2 \left(\epsilon _2\right){}^+}{3 \sqrt{3}}+\frac{2}{3} & -\frac{2 \left(\epsilon _3\right){}^+}{3 \sqrt{3}} & -\frac{\left(\epsilon _4\right){}^+}{3
   \sqrt{6}}-\frac{1}{3 \sqrt{2}} & \frac{\left(\epsilon _5\right){}^+}{3 \sqrt{6}} & -\frac{\left(\epsilon _6\right){}^+}{\sqrt{10}} & 0 & 0 & 0 & 0 & 0 & 0 \\
 \frac{\left(\epsilon _1\right){}^+}{3}+\frac{1}{\sqrt{3}} & -\frac{\left(\epsilon _2\right){}^+}{3 \sqrt{3}}-\frac{1}{3} & -\frac{2 \left(\epsilon _3\right){}^+}{3
   \sqrt{3}} & -\frac{\left(\epsilon _4\right){}^+}{3 \sqrt{6}}-\frac{1}{3 \sqrt{2}} & -\frac{1}{3} \sqrt{\frac{2}{3}} \left(\epsilon _5\right){}^+ & \frac{1}{3}
   \sqrt{\frac{2}{5}} \left(\epsilon _6\right){}^+ & 0 & 0 & 0 & 0 & 0 & 0 \\
 0 & 0 & 0 & 0 & 0 & 0 & -\frac{\left(\epsilon _1\right){}^-}{3 \sqrt{2}}-\frac{1}{\sqrt{6}} & -\frac{\left(\epsilon _2\right){}^-}{3 \sqrt{6}}-\frac{1}{3 \sqrt{2}} &
   -\frac{1}{3} \sqrt{\frac{2}{3}} \left(\epsilon _3\right){}^- & \frac{\left(\epsilon _4\right){}^-}{3 \sqrt{3}}+\frac{1}{3} & \frac{2 \left(\epsilon _5\right){}^-}{3
   \sqrt{3}} & \frac{2 \left(\epsilon _6\right){}^-}{3 \sqrt{5}} \\
 -\frac{\left(\epsilon _1\right){}^+}{3 \sqrt{2}}-\frac{1}{\sqrt{6}} & -\frac{\left(\epsilon _2\right){}^+}{3 \sqrt{6}}-\frac{1}{3 \sqrt{2}} & -\frac{1}{3}
   \sqrt{\frac{2}{3}} \left(\epsilon _3\right){}^+ & \frac{\left(\epsilon _4\right){}^+}{3 \sqrt{3}}+\frac{1}{3} & \frac{2 \left(\epsilon _5\right){}^+}{3 \sqrt{3}} &
   \frac{2 \left(\epsilon _6\right){}^+}{3 \sqrt{5}} & 0 & 0 & 0 & 0 & 0 & 0 \\
 0 & 0 & 0 & 0 & 0 & 0 & \frac{\left(\epsilon _1\right){}^-}{3}+\frac{1}{\sqrt{3}} & -\frac{\left(\epsilon _2\right){}^-}{3 \sqrt{3}}-\frac{1}{3} & -\frac{2
   \left(\epsilon _3\right){}^-}{3 \sqrt{3}} & -\frac{\left(\epsilon _4\right){}^-}{3 \sqrt{6}}-\frac{1}{3 \sqrt{2}} & -\frac{1}{3} \sqrt{\frac{2}{3}} \left(\epsilon
   _5\right){}^- & \frac{1}{3} \sqrt{\frac{2}{5}} \left(\epsilon _6\right){}^- \\
 0 & 0 & 0 & -\frac{\left(\epsilon _4\right){}^+}{\sqrt{6}}-\frac{1}{\sqrt{2}} & \frac{\left(\epsilon _5\right){}^+}{\sqrt{6}} & \frac{\left(\epsilon
   _6\right){}^+}{\sqrt{10}} & 0 & 0 & 0 & 0 & 0 & 0 \\
 0 & 0 & 0 & 0 & 0 & 0 & 0 & \frac{1}{3} \sqrt{\frac{2}{3}} \left(\epsilon _2\right){}^-+\frac{\sqrt{2}}{3} & -\frac{1}{3} \sqrt{\frac{2}{3}} \left(\epsilon
   _3\right){}^- & \frac{\left(\epsilon _4\right){}^-}{3 \sqrt{3}}+\frac{1}{3} & -\frac{\left(\epsilon _5\right){}^-}{3 \sqrt{3}} & \frac{\left(\epsilon
   _6\right){}^-}{\sqrt{5}} \\
 0 & \frac{1}{3} \sqrt{\frac{2}{3}} \left(\epsilon _2\right){}^++\frac{\sqrt{2}}{3} & -\frac{1}{3} \sqrt{\frac{2}{3}} \left(\epsilon _3\right){}^+ &
   \frac{\left(\epsilon _4\right){}^+}{3 \sqrt{3}}+\frac{1}{3} & -\frac{\left(\epsilon _5\right){}^+}{3 \sqrt{3}} & \frac{\left(\epsilon _6\right){}^+}{\sqrt{5}} & 0 &
   0 & 0 & 0 & 0 & 0 \\
 0 & 0 & 0 & 0 & 0 & 0 & 0 & 0 & 0 & -\frac{\left(\epsilon _4\right){}^-}{\sqrt{6}}-\frac{1}{\sqrt{2}} & \frac{\left(\epsilon _5\right){}^-}{\sqrt{6}} &
   \frac{\left(\epsilon _6\right){}^-}{\sqrt{10}} \\
 \frac{\left(\epsilon _1\right){}^+}{3 \sqrt{2}}-\frac{1}{\sqrt{6}} & \frac{1}{3 \sqrt{2}}-\frac{\left(\epsilon _2\right){}^+}{3 \sqrt{6}} & -\frac{1}{3}
   \sqrt{\frac{2}{3}} \left(\epsilon _3\right){}^+ & \frac{\left(\epsilon _4\right){}^+}{3 \sqrt{3}}-\frac{1}{3} & \frac{2 \left(\epsilon _5\right){}^+}{3 \sqrt{3}} &
   -\frac{2 \left(\epsilon _6\right){}^+}{3 \sqrt{5}} & 0 & 0 & 0 & 0 & 0 & 0 \\
 0 & 0 & 0 & 0 & 0 & 0 & \frac{1}{\sqrt{3}}-\frac{\left(\epsilon _1\right){}^-}{3} & \frac{1}{3}-\frac{\left(\epsilon _2\right){}^-}{3 \sqrt{3}} & -\frac{2
   \left(\epsilon _3\right){}^-}{3 \sqrt{3}} & \frac{1}{3 \sqrt{2}}-\frac{\left(\epsilon _4\right){}^-}{3 \sqrt{6}} & -\frac{1}{3} \sqrt{\frac{2}{3}} \left(\epsilon
   _5\right){}^- & -\frac{1}{\sqrt{15}} \\
 \frac{1}{\sqrt{3}}-\frac{\left(\epsilon _1\right){}^+}{3} & \frac{1}{3}-\frac{\left(\epsilon _2\right){}^+}{3 \sqrt{3}} & -\frac{2 \left(\epsilon _3\right){}^+}{3
   \sqrt{3}} & \frac{1}{3 \sqrt{2}}-\frac{\left(\epsilon _4\right){}^+}{3 \sqrt{6}} & -\frac{1}{3} \sqrt{\frac{2}{3}} \left(\epsilon _5\right){}^+ & -\frac{1}{3}
   \sqrt{\frac{2}{5}} \left(\epsilon _6\right){}^+ & 0 & 0 & 0 & 0 & 0 & 0 \\
 0 & 0 & 0 & 0 & 0 & 0 & \frac{\left(\epsilon _1\right){}^-}{3 \sqrt{2}}-\frac{1}{\sqrt{6}} & \frac{1}{3 \sqrt{2}}-\frac{\left(\epsilon _2\right){}^-}{3 \sqrt{6}} &
   -\frac{1}{3} \sqrt{\frac{2}{3}} \left(\epsilon _3\right){}^- & \frac{\left(\epsilon _4\right){}^-}{3 \sqrt{3}}-\frac{1}{3} & \frac{2 \left(\epsilon _5\right){}^-}{3
   \sqrt{3}} & -\frac{2 \left(\epsilon _6\right){}^-}{3 \sqrt{5}} \\
 0 & 0 & 0 & 0 & 0 & 0 & 0 & \frac{2 \left(\epsilon _2\right){}^-}{3 \sqrt{3}}-\frac{2}{3} & -\frac{2 \left(\epsilon _3\right){}^-}{3 \sqrt{3}} & \frac{1}{3
   \sqrt{2}}-\frac{\left(\epsilon _4\right){}^-}{3 \sqrt{6}} & \frac{\left(\epsilon _5\right){}^-}{3 \sqrt{6}} & -\frac{\left(\epsilon _6\right){}^-}{\sqrt{10}} \\
 0 & \frac{2 \left(\epsilon _2\right){}^+}{3 \sqrt{3}}-\frac{2}{3} & -\frac{2 \left(\epsilon _3\right){}^+}{3 \sqrt{3}} & \frac{1}{3 \sqrt{2}}-\frac{\left(\epsilon
   _4\right){}^+}{3 \sqrt{6}} & \frac{\left(\epsilon _5\right){}^+}{3 \sqrt{6}} & -\frac{\left(\epsilon _6\right){}^+}{\sqrt{10}} & 0 & 0 & 0 & 0 & 0 & 0 \\
 \frac{\left(\epsilon _1\right){}^+}{3}-\frac{1}{\sqrt{3}} & \frac{1}{3}-\frac{\left(\epsilon _2\right){}^+}{3 \sqrt{3}} & -\frac{2 \left(\epsilon _3\right){}^+}{3
   \sqrt{3}} & \frac{1}{3 \sqrt{2}}-\frac{\left(\epsilon _4\right){}^+}{3 \sqrt{6}} & -\frac{1}{3} \sqrt{\frac{2}{3}} \left(\epsilon _5\right){}^+ & \frac{1}{3}
   \sqrt{\frac{2}{5}} \left(\epsilon _6\right){}^+ & 0 & 0 & 0 & 0 & 0 & 0 \\
 0 & 0 & 0 & 0 & 0 & 0 & \frac{1}{\sqrt{6}}-\frac{\left(\epsilon _1\right){}^-}{3 \sqrt{2}} & \frac{1}{3 \sqrt{2}}-\frac{\left(\epsilon _2\right){}^-}{3 \sqrt{6}} &
   -\frac{1}{3} \sqrt{\frac{2}{3}} \left(\epsilon _3\right){}^- & \frac{\left(\epsilon _4\right){}^-}{3 \sqrt{3}}-\frac{1}{3} & \frac{2 \left(\epsilon _5\right){}^-}{3
   \sqrt{3}} & \frac{2 \left(\epsilon _6\right){}^-}{3 \sqrt{5}} \\
 \frac{1}{\sqrt{6}}-\frac{\left(\epsilon _1\right){}^+}{3 \sqrt{2}} & \frac{1}{3 \sqrt{2}}-\frac{\left(\epsilon _2\right){}^+}{3 \sqrt{6}} & -\frac{1}{3}
   \sqrt{\frac{2}{3}} \left(\epsilon _3\right){}^+ & \frac{\left(\epsilon _4\right){}^+}{3 \sqrt{3}}-\frac{1}{3} & \frac{2 \left(\epsilon _5\right){}^+}{3 \sqrt{3}} &
   \frac{2 \left(\epsilon _6\right){}^+}{3 \sqrt{5}} & 0 & 0 & 0 & 0 & 0 & 0 \\
 0 & 0 & 0 & 0 & 0 & 0 & \frac{\left(\epsilon _1\right){}^-}{3}-\frac{1}{\sqrt{3}} & \frac{1}{3}-\frac{\left(\epsilon _2\right){}^-}{3 \sqrt{3}} & -\frac{2
   \left(\epsilon _3\right){}^-}{3 \sqrt{3}} & \frac{1}{3 \sqrt{2}}-\frac{\left(\epsilon _4\right){}^-}{3 \sqrt{6}} & -\frac{1}{3} \sqrt{\frac{2}{3}} \left(\epsilon
   _5\right){}^- & \frac{1}{3} \sqrt{\frac{2}{5}} \left(\epsilon _6\right){}^- \\
 0 & 0 & 0 & \frac{1}{\sqrt{2}}-\frac{\left(\epsilon _4\right){}^+}{\sqrt{6}} & \frac{\left(\epsilon _5\right){}^+}{\sqrt{6}} & \frac{\left(\epsilon
   _6\right){}^+}{\sqrt{10}} & 0 & 0 & 0 & 0 & 0 & 0 \\
 0 & 0 & 0 & 0 & 0 & 0 & 0 & \frac{1}{3} \sqrt{\frac{2}{3}} \left(\epsilon _2\right){}^--\frac{\sqrt{2}}{3} & -\frac{1}{3} \sqrt{\frac{2}{3}} \left(\epsilon
   _3\right){}^- & \frac{\left(\epsilon _4\right){}^-}{3 \sqrt{3}}-\frac{1}{3} & -\frac{\left(\epsilon _5\right){}^-}{3 \sqrt{3}} & \frac{\left(\epsilon
   _6\right){}^-}{\sqrt{5}} \\
 0 & \frac{1}{3} \sqrt{\frac{2}{3}} \left(\epsilon _2\right){}^+-\frac{\sqrt{2}}{3} & -\frac{1}{3} \sqrt{\frac{2}{3}} \left(\epsilon _3\right){}^+ &
   \frac{\left(\epsilon _4\right){}^+}{3 \sqrt{3}}-\frac{1}{3} & -\frac{\left(\epsilon _5\right){}^+}{3 \sqrt{3}} & \frac{\left(\epsilon _6\right){}^+}{\sqrt{5}} & 0 &
   0 & 0 & 0 & 0 & 0 \\
 0 & 0 & 0 & 0 & 0 & 0 & 0 & 0 & 0 & \frac{1}{\sqrt{2}}-\frac{\left(\epsilon _4\right){}^-}{\sqrt{6}} & \frac{\left(\epsilon _5\right){}^-}{\sqrt{6}} &
   \frac{\left(\epsilon _6\right){}^-}{\sqrt{10}} \\
\end{array}
\right)
\begin{pmatrix}
    A^{\frac{1}{ 2},0,+}(\frac{1}{2})\\
     A^{\frac{1}{ 2},1,+}(\frac{1}{2})\\
      A^{\frac{1}{ 2},1,+}(\frac{3}{2})\\
       A^{\frac{3}{ 2},1,+}(\frac{1}{2})\\
        A^{\frac{3}{ 2},1,+}(\frac{3}{2})\\
         A^{\frac{3}{ 2},2,+}(\frac{3}{2})\\
           A^{\frac{1}{ 2},0,+}(\frac{1}{2})\\
     A^{\frac{1}{ 2},1,-}(\frac{1}{2})\\
      A^{\frac{1}{ 2},1,-}(\frac{3}{2})\\
       A^{\frac{3}{ 2},1,-}(\frac{1}{2})\\
        A^{\frac{3}{ 2},1,-}(\frac{3}{2})\\
         A^{\frac{3}{ 2},2,-}(\frac{3}{2})\\
\end{pmatrix}
$
}
$
}
\end{align}

\end{appendix}
\bibliographystyle{JHEP}
\bibliography{baryonCPV}
\end{document}